\documentstyle[proceedings,numreferences]{crckapb}

\newcommand{\Nr}{{{\bf r}}}
\newcommand{\qq}{\begin{eqnarray}}

\newcommand{\ee}{{\rm e}}

\newcommand{\qqq}{\end{eqnarray}}

\newcommand{\CC}{{\cal C}}
\newcommand{\CD}{{\cal D}}

\newcommand{\CO}{{\cal O}}

\newcommand{\CT}{{\cal T}}

\newcommand{\s}{\hspace{0.05cm}}
\newcommand{\m}{\hspace{0.025cm}}

\newcommand{\unr}{{\underline{\bf r}}}

\input{psfig}
\begin{opening}
\title{
  INTERMITTENCY OF PASSIVE ADVECTION$^\dagger$
}
\runningtitle{INTERMITTENCY OF PASSIVE ADVECTION}

\author{KRZYSZTOF GAWEDZKI}
\institute{CNRS, IHES, 91440 Bures-sur-Yvette, France}
\end{opening}
\runningauthor{K. GAWEDZKI}
\begin{document}

\section{Introduction}

An explanation of the origin of intermittency\footnote{$\hspace{-0.63cm}
^{\dagger}$ contribution to the proceedings of the $7^{\rm th}$ European 
Turbulence Conference, Saint Jean \hspace*{0.2cm}Cap Ferrat, 1998}
in the fully developed turbulence remains one 
of the main open problems of theoretical hydrodynamics. 
Quantitatively, intermittency is measured by deviations 
from Kolmogorov's scaling of the velocity correlators. 
Below, we shall describe a recent progress 
in the understanding of intermittency in a simple system describing 
the passive advection of a scalar quantity (temperature, 
density of a pollutant) by a random velocity field. The system, 
maintained in a stationary state by a large scale source, 
exhibits a down-scale energy cascade. If the molecular diffusion 
is small, an intermittent inertial range sets in \cite{Anton}. 
We shall explain this phenomenon by identifying the origin 
of the anomalous scaling of scalar correlators in a simple 
model of passive advection due to Kraichnan \cite{Kr68}.

\section{Kraichnan model}


The advection of a scalar quantity $\theta(t,{\bf r})$ 
by a velocity field ${\bf v}(t,{\bf r})$ is described 
by the linear differential equation
\qq
\partial_t\theta\,+\,{\bf v}\cdot{\bf\nabla}\theta\,
=\, f
\label{ps}
\qqq
where $f(t,{\bf r})$ represents the external source of the scalar. 
If the source vanishes then the scalar 
is carried along by the flow:
\qq
\theta(t,{\bf r}(t))\,=\,\theta(t_0,{\bf r}_0)
\label{homo}
\qqq
where ${\bf r}(t)$ is the (Lagrangian) trajectory of the fluid 
particle located at time $t_0$ at ${\bf r}_0$: 
\qq
{d{\bf r}\over dt}\,=\, {\bf v}(t,{\bf r})\,,
\qquad {\bf r}(t_0)={\bf r}_0\,.
\label{lagr}
\qqq
The non-zero source $f$ keeps creating the scalar along 
the Lagrangian trajectory and Eq.\,\,(\ref{homo}) is modified to
\qq
\theta(t,{\bf r}(t))\,=\,\theta(t_0,{\bf r}_0)\,+\,
\int_{_{t_0}}^{^t}\hspace{-0.2cm}ds\ f(s,{\bf r}(s))\,.
\label{inhomo}
\qqq
In the presence of diffusion of the scalar the above 
equations require small changes:
\vskip 0.2cm

\noindent 1. \ Eq.\,\,(\ref{ps}) picks up a term 
$\kappa{\bf\nabla}^2\theta$ on the right hand side with $\kappa$
standing for the diffusion constant,
\vskip 0.1cm

\noindent 2. \ Brownian motions $\beta(t)$ should be 
superimposed on the Lagrangian trajectories 
by adding the term $\kappa\m{d\beta\over dt}$ to the velocity 
in Eq.\,\,(\ref{lagr}), 
\vskip 0.1cm

\noindent 3. \ the right hand sides of Eqs.\,\,(\ref{homo}) and 
(\ref{inhomo}) should be averaged over $\beta$.
\vskip 0.3cm

The Kraichnan model of the passive advection of the scalar 
\cite{Kr68} assumes velocities decorrelated at different times,
and, for a fixed time, distributed as a Gaussian (i.e.
non-intermittent) field with mean zero and a non-smooth 
typical behavior in space
\qq
\vert {\bf v}(t,{\bf r})-{\bf v}(t,{\bf r}')\vert\ \sim\ 
\vert{\bf r}-{\bf r}'\vert^{\xi/2}\,
\label{Hold}
\qqq
where $\xi$ is a fixed parameter between $0$ and $2$. 
This may be achieved by imposing the velocity covariance 
\qq
\left\langle {\bf v}^\alpha(t,{\bf r})\,\, {\bf v}^\beta(t',
{\bf r}')\right\rangle\ =\ 
\delta(t-t')\ \CD^{\alpha\beta}({\bf r}-{\bf r}')
\label{cov}
\qqq
with 
\qq
\CD^{\alpha\beta}(0)-\CD^{\alpha\beta}({\bf r})\ =\ 
D_0\, r^\xi\left[(d-1+\xi)\,\delta^{\alpha\beta}\,
-\,\xi\,{_{r^\alpha\m r^\beta}\over^{r^2}}
\right]\ \equiv\ D^{\alpha\beta}({\bf r})
\qqq
where $d$ denotes the space dimension. The typical behavior 
(\ref{Hold}) follows since $D^{\alpha\beta}({\bf r})\propto r^\xi$. 
The incompressibility ${\bf\nabla}\cdot{\bf v}=0$ of the velocity 
field is assured by the relation $\partial_{r^\alpha}
\CD^{\alpha\beta}({\bf r})=0$. 
\vskip 0.3cm

We shall be interested in the effect caused by a steady
injection of the scalar on distances longer than
some large scale $L$. One may conveniently model such
a source by a random Gaussian field $f(t,\Nr)$ with mean zero
and covariance
\qq
\left\langle f(t,{\bf r})\,\, f(t',{\bf r}')
\right\rangle\ =\ 
\delta(t-t')\ \CC({_{{\bf r}-{\bf r}'}\over^L})
\label{cov1}
\qqq
where $\CC({{\bf r}\over L})$ 
is approximately constant for $r\ll L$
and decays fast for $r\gg L$. We shall assume $f$ independent 
of the velocities ${\bf v}$. 

\section{Steady state}

The evolution equations of the hydrodynamical type  
imply identities for the correlators of the evolving 
quantities known under the name of Hopf equations. 
These equation usually couple the correlators
with different number of points. The case
of the passive advection is no exception since its
Hopf equations couple the scalar equal-time correlators 
\qq
F_N(t;\unr)\ =\ \Big\langle\prod\limits_{n=1}^N\theta(t,{\bf r}_n) 
\Big\rangle
\label{Npoint}
\qqq
to the mixed correlators involving both the scalar and the
velocity field. However, in the Kraichnan model with temporarily
decorrelated velocities, the mixed correlators may
be expressed by the known $2$-point function of the
velocities and the correlators of the scalar alone.
The resulting Hopf equations may be easily seen to take the form
\qq
\partial_t\m F_N\ =\ -\m M_N F_N\ +\ \CC\otimes F_{N-2}
\label{diffe}
\qqq
with the shorthand notation
\qq
(\CC\otimes F_{N-2})(t;\unr)\ =\ \sum\limits_{n<m}
\CC({_{\Nr_n-\Nr_m}\over^L})\ F_{N-2}(t;\m\Nr_1,
\mathop{......}\limits_{\hat{n}\ \hat{m}}\,,\Nr_N)
\qqq
and with
\qq
M_N\ =\ 
-\,{_1\over^2}\sum\limits_{n,m
\atop\alpha\beta}\CD^{\alpha\beta}({\bf r}_n-{\bf r}_m)\,
\,\partial_{r_n^\alpha}\partial_{r_m^\beta}
\label{MN}
\qqq
being a symmetric (due to the incompressibility), positive, 
singular elliptic differential operator of the $2^{\rm nd}$ 
order. In the presence of the diffusion, $M_N$ should be modified 
by the subtraction of $\sum_n\kappa({\bf\nabla}_{{\bf r}_n})^2$.
In the translationally invariant sector, i.e. for the 
homogeneous distributions of the scalar, the matrix
$\CD^{\alpha\beta}(\Nr)$ in the definition of $M_N$ may
be replaced by $-\m D^{\alpha\beta}(\Nr)$ since
the contribution of $\CD^{\alpha\beta}(0)$ drops
out. Below, we shall use this scaling form of dimension 
$(length)^{\xi-2}$ of the operators $M_N$ assuming also
that the diffusion constant $\kappa$ has been taken to zero.
\vskip 0.3cm

The Hopf equations (\ref{diffe}) may be solved by induction 
with respect to $N$. Denoting by $P_N(t;\unr,\unr')$ 
the Green functions $\ee^{-t\m M_N}(\unr,\unr')$, we obtain
\qq
F_N(t;\unr)&=&\int P_N(t-t_0;\m\unr,\unr')
\,\, F_N(t_0;\m \unr')\,\, d\unr'\cr
&+&\int_{_{t_0}}^{^{t}}\hspace{-0.2cm}ds\int P_N(t-s;\m\unr,\unr')\ 
(\CC\otimes F_{N-2})(s,\unr')\,\ d\unr'\,.
\label{nhsol}
\qqq
For a concentrated initial distribution of $\theta$ with
fast decaying $N$-point functions and for $t\rightarrow\infty$, 
the $N$-point functions $F_N(t;\unr)$
tend to the solution $F_N(\unr)$ of the stationary version
of the Hopf equations
\qq
M_N F_N\ =\ \CC\otimes F_{N-2}
\label{stat}
\qqq
inductively determined by the relations
\qq
F_N(\unr)\ =\  
\int G_N(\unr,\unr')\ (\CC\otimes F_{N-2})(\unr')\,\, d\unr'\,.
\label{sol}
\qqq
where $\,G_N(\unr,\unr')\,=\,\int_{_0}^{^\infty}\hspace{-0.3cm}ds\ 
P_N(s;\m\unr,\unr')\,$ is the kernel of the inverse of the operator 
$M_N$. In particular, the limiting stationary distribution 
is independent of the initial one and has vanishing odd-point 
functions. 

\section{Zero mode dominance}

We are interested in the scaling properties of the stationary 
$N$-point functions $F_N(\unr)$ for the injection scale $L$ 
large w.r.t. point separations, i.e. in the inertial range. 
Since $M_N$ have the scaling dimension of $(length)^{\xi-2}$
and $\CC({\Nr\over L})$ is approximately
constant for large $L$, we might expect that the solutions 
of the chain (\ref{stat}) of equations scale like
\qq
F_N(\lambda\m\unr)\ =\ \lambda^{(2-\xi)N\over 2}\,\, F_N(\unr)\,.
\qqq
This would be the normal Kolmogorov-Obukhov-Corrsin scaling.
\vskip 0.3cm

In the 1994 paper \cite{Kr94}, Kraichnan has argued in favor 
of the anomalous scaling of the scalar correlators.
His paper steered a renewed interest in the problem which
led to the discovery of a simple mechanism by which
the correlators avoid the normal scaling. As was realized in 
\cite{GK} and \cite{CFKL}, see also \cite{SS}, for large $L$ 
the normally scaling contributions to $F_N$ are dominated 
by the ones of the {\it zero modes} of the operators $M_N$. 
The stationary Hopf equations (\ref{stat}) leave the freedom
to add the solutions of their homogeneous version
$M_N f_N=0$. It is not obvious, however, whether the solution 
(\ref{sol}), which is defined unambiguously, contains the scaling 
zero modes which become leading in the inertial range. 
Such zero mode contributions were, indeed, found 
in the 2-point function. The latter may be calculated exactly 
\cite{Kr68} with the result
\qq
F_2(r)\ =\ A_0\, L^{2-\xi}\ +\ A_1 r^{2-\xi}\ +\ ...
\qqq
where the terms vanishing as inverse powers when 
$L\rightarrow\infty$ were omitted. The constant $\propto L^{2-\xi}$ 
is a zero mode of operator $M_2$ and it dominates for $r\ll L$. 
The constant terms, as well as the ones independent 
of some of ${\bf r}_n$'s, however, do not contribute 
to the structure functions
\qq
S_N(r)\ =\ \left\langle\,(\theta(\Nr)-\theta({\bf 0}))^N
\right\rangle
\label{strf}
\qqq
of the scalar. As a result, the $2$-point structure function
scales as $r^{2-\xi}$, i.e. normally.
\vskip 0.3cm

What about the higher-point functions?
In \cite{GK} and in \cite{CFKL} it was shown that the $4$-point 
structure function is dominated for $r\ll L$ by the contribution 
of a zero mode of the operator $M_4$ for, respectively, small 
velocity exponent $\xi$ and large space dimension $d$. In \cite{BGK} 
and \cite{CF} these results have been extended to $N$-point
functions with the result
\qq
S_{N}(r)\ \ \propto\ \ r^{\zeta_N}\qquad{\rm with}\quad
\zeta_N\ =\ {_{(2-\xi)N}\over^2}\ -\ \cases{{{N(N-2)\m\xi}\over{
2\m(d+2)}}\ +\ \CO(\xi^2)\,,\cr\cr{{N(N-2)\m\xi}\over{2\m d}}\ 
+\ \CO(d^{-2})\,.}
\label{ascal}
\qqq
The scaling dimensions of the relevant zero modes where
found by restricting operators $M_N$ to scaling functions. 
Upon such a restriction, $M_N$'s become operators with discrete 
spectrum to which standard perturbative technics apply.
A more difficult perturbative analysis around $\xi=2$
\cite{SS}\cite{GLPP}\cite{BCFL} confirms the zero
mode dominance of the inertial-range scaling also in this
regime.

\section{Slow modes}

The Green functions $P_N(t;\unr,\unr')=\ee^{-t\, M_N}(\unr,\unr')$
have a natural interpretation in the language of 
Lagrangian trajectories. They are the joint probability 
distribution functions (PDF's) of the time $t$ positions 
$\Nr_n$ of the trajectories $\Nr_n(s)$ starting at time zero 
at points $\Nr_n'$. This follows from the $\CC=0$ version 
of Eq.\,\,(\ref{nhsol}) if we recall that, in the absence 
of the source, the scalar density is carried along 
the Lagrangian trajectories. This interpretation 
of $P_N(t;\unr,\unr')$ shows that, effectively, the (differences
of the) Lagrangian trajectories undergo a simple diffusion 
process with the space dependent diffusion matrix ${_1\over^2}
D^{\alpha\beta}(\Nr_n-\Nr_m)\,\propto\,\vert{\bf r}_n-{\bf r}_m
\vert^\xi$ so that they diffuse very slowly when they are 
close but faster and faster when, eventually, they separate.
\vskip 0.3cm

Let us look closer at the effective diffusion
of the Lagrangian trajectories. The scaling property 
of operators $M_N$ implies that
\qq
P_N(\lambda^{2-\xi}\m t;\m\lambda\unr,\lambda\unr_0)
\ =\ \lambda^{-Nd}\, P_N(t;\unr,\unr_0)\,.
\label{sc}
\qqq
Thus the {\it distances} scale as $(time)^{^{1\over 2-\xi}}$
as compared to $(time)^{^{1\over 2}}$ in the standard diffusion.
We may probe the effective spread of Lagrangian trajectories
with scaling functions $f_N$ such that $\,f_N(\lambda\unr)=
\lambda^{\sigma}f_N(\unr)\m$, \, e.g. with the sum of squares 
of distances between ${\bf r}_n$'s. The average of $f_N$ 
over the time zero positions of $N$ Lagrangian trajectories 
which at time $t$ pass by points $\unr$ 
is$^1$\footnote{\hspace{-0.3cm}$^1$ for the later convenience, 
we study the backward evolution in time}
\qq
\langle\s f_N\m\rangle_{_{t,\m\unr}}\hspace{-0.2cm}&=&
\hspace{-0.2cm}\int P_N(t;\unr,\unr')\,\, f_N(\unr')\,\, d\unr' 
\,=\,\lambda^{-\sigma}\hspace{-0.1cm}
\int\hspace{-0.1cm} P_N(\lambda^{2-\xi}\m t;\m\lambda\m\unr,\unr')
\,\,f_N(\unr')\,\, d\unr'\cr
&=&\hspace{-0.1cm}({_t\over^\tau})^{\sigma\over 2-\xi}
\int P_N(\tau;\m({_\tau\over^t}
)^{{1\over 2-\xi}}\unr\m,\,\unr')\,\, f_N(\unr')
\,\, d\unr' 
\label{avb}
\qqq
where we have used the scaling properties of $f_N$
and $P_N$ and have set $\tau=\lambda^{2-\xi}t$. 
If we let $t$ grow and keep $\tau$ constant, the last 
integral may be expected to tend to the limit $\,\int 
P_N(\tau;\m 0,\unr')\, f_N(\unr')\, d\unr'\,$ so that, 
if the latter integral does not vanish, 
\qq
\langle\s f_N\m\rangle_{_{t,\m\unr}}\ \ \,\propto\ \ \ 
({_t\over^\tau})^{\sigma\over 2-\xi}\,
\label{typb}
\qqq
This is the super-diffusive behavior which sets in for typical
scaling probes $f_{N}$. Suppose, however, that $f_N$
is a scaling zero mode of $M_N$. Since
\qq
{d\over dt}\,\langle\s f_N\m\rangle_{_{t,\m\unr}}
\ =\ -\int P_N(t;\unr,\unr')\ (M_N f_N)(\unr')\,\, d\unr' 
\ =\ 0\,,
\qqq
the average $\langle\s f_N\m\rangle_{t,\m\unr}$ is  
constant in time instead of growing like $({t\over\tau})
^{\sigma\over 2-\xi}$: the zero modes describe 
structures preserved in mean by the Lagrangian flow.
\vskip 0.3cm

A constant or linear functions are the obvious 
examples of scaling zero modes but a closer analysis 
\cite{slow} of operators $M_N$ shows that there is 
a whole discrete series $f_{N,i}$ of them 
with scaling dimensions $\sigma_i\geq 0$. 
Besides, each such scaling zero mode 
gives rise to a tower of {\it slow modes} $f_{N,ik}$, 
$k=0,1,\dots,$ starting at $f_{N,i0}=f_{N,i}$,
with scaling dimensions $\sigma_i+(2-\xi)k$. The averages 
$\langle\s f_{N,ik} \rangle_{t,\unr}$ are polynomials 
of order $k$ in $t$ so that they grow slower than the typical 
behavior (\ref{typb}). All these scaling modes appear 
in the small $\lambda$ asymptotic expansion \cite{slow}
\qq
P_N(\tau;\lambda\unr,\unr')\ =\ \sum\limits_{i,k}\,
\lambda^{\sigma_i+(2-\xi)k}\,\, 
f_{N,ik}(\unr)\ g_{N,ik}(\tau;\unr') 
\label{asex}
\qqq
which, when inserted on the right hand
side of Eq.\s\s(\ref{avb}), gives
\qq
\langle\s f_N\m\rangle_{_{t,\m\unr}}\ =\ 
\sum_{i,k}\,\,({_t\over^\tau})^{{\sigma-\sigma_i
\over 2-\xi}\m-\m k}\,\, f_{N,ik}(\unr)\ 
\int g_{N,ik}(\tau;\unr')\,\,f_N(\unr')\,\, 
d\unr'\,.
\label{asex1}
\qqq
The functions $g_{N,ik}(\tau;\unr)$ are finite and decay fast 
for large $\unr$. For generic $f_N$, the constant zero 
mode $f_{N,00}$ corresponding to $g_{N,00}(\tau;\unr')
=P_N(\tau;{\bf 0},\unr')$ gives the dominant
contribution leading to the behavior
(\ref{typb}). This term (and many others) vanishes
for $f_N=f_{N,jl}$ resulting in the slower growth.
\vskip 0.3cm

Upon integration over $t$, the asymptotic expansion 
(\ref{asex}) induces the one for the kernels 
of the inverses of $M_N$'s:
\qq
G_N(\lambda\unr,\unr')\ =\ \sum\limits_{i}\m \lambda^{\sigma_i}
\,\m f_{N,i}(\unr)\ h_{N,i}(\unr')
\qqq
(the contributions of the slow modes disappear under the 
time integration). This expansion is behind the zero-mode
dominance of the short-distance behavior of the iterative 
solutions (\ref{sol}) for the stationary $N$-point functions
of the scalar. The subtractions required in the passage 
from the correlation functions to the structure functions 
leave only the contributions of the zero modes $f_{N,i}$
fully dependent on all $\Nr_n$'s.

\section{Fuzzy trajectories}

The asymptotic expansion (\ref{asex}) describes what
happens when the final points of the Lagrangian trajectories 
are taken together. It was established with mathematical 
rigor for the case of two Lagrangian trajectories 
and was checked in perturbative approaches for $N>2$. 
At the first glance, it may look bizarre. 
Indeed, we may naively expect that if the final points 
${\bf r}_{n}$ of the trajectories converge then their initial 
points should also do with the joint PDF $\, P_N(t;\m\lambda
\m\unr,\m\unr')\,$ becoming proportional to \,$\prod\delta
({\bf r}'_n-{\bf r}'_{n+1})\,$ when $\lambda\rightarrow 0$. 
Instead, $\, P_N(t;\m\lambda\m\unr,\m\unr')\,$ tends 
to a regular limit $\, g_{N,00}(t;\unr')$.
The mechanism by which $P_N(t;\unr,\unr')$ avoids the singularity 
at $\unr=0$ is somewhat subtle \cite{slow}. Recall that 
the Lagrangian trajectories satisfy the differential equation
(\ref{lagr}). The uniqueness of solutions of such 
an equation requires the Lipschitz condition $|{\bf v}(t,\Nr)-
{\bf v}(t,\Nr')|\sim|\Nr-\Nr'|$ for $\Nr'$ tending to $\Nr$. 
But our velocities are only H\"{o}lder continuous in $\Nr$ 
with exponent ${\xi\over 2}<1$, see Eq.\,\,(\ref{Hold}). 
The resulting non-uniqueness of the Lagrangian trajectories 
passing through a fixed point violates the Newton-Leibniz 
paradigm and allows for a continuum of trajectories 
with coincident final points. Although the trajectories 
keep collapsing continuously, the probabilistic quantities 
such as the joint PDF's $P_N(t;\unr,\unr')$ still make 
perfect sense. They reflect, however, the fuzzyness 
of the trajectories in their asymptotics (\ref{asex}).
\vskip 0.3cm

A less rigorous-minded reader might object that the non-uniqueness
of the Lagrangian trajectories is a mathematical nuisance since
more realistic turbulent velocities, even when showing 
the behavior (\ref{Hold}) with $\xi$ close to ${2\over 3}$
in the inertial range, are smoothed on short distances 
by the viscous effects so that their Lagrangian trajectories 
are unique. A closer examination, shows, however, that such
Lagrangian trajectories, although uniquely determined by sharp 
values at one time, exhibit a {\it sensitive dependence} on 
these values signaled by non-zero {\it Lyapunov exponents}. 
As a result, if the fixed-time positions of the trajectories 
have an $\epsilon$-spread, the trajectories at different times 
are spread in a large region and will stay such if the viscosity 
is taken to zero first and the spread $\epsilon$ 
only next. The non-uniqueness of the trajectories 
caused by their continuous collapse is simply a useful 
mathematical abstraction describing a real physical phenomenon: 
a fast spread of the trajectories in each realization of 
the high Reynolds number velocity field. 
Such a spread makes the identification of the trajectories 
practically impossible. It should be remarked that 
in the incompressible case, the collapse of the trajectories 
must be accompanied symmetrically by their branching since 
$P_N(t;\unr,\unr')=P_N(t;\unr',\unr)$. In compressible flows, 
this symmetry is broken. In particular, in the utmost 
compressible Burgers flows, the trajectories only collapse 
together (in a discrete process) sticking to the shocks.

\section{Numerical results}

The region of $\xi$ neither close to $0$ nor to $2$ has 
up to now defied an exact analysis. The first numerical 
simulations of the system \cite{KYC}\cite{FGLP} were based
on the direct solution of the scalar equation (\ref{ps}).
They did not cover the region of small $\xi$ and did not 
permit to discriminate between the small ($\propto \xi$) 
anomalous structure-function exponents implied 
by the perturbative zero-mode analysis and the different
$(\propto 1)$ predictions of a closure 
of the structure-function equations proposed in \cite{Kr94}, 
see also \cite{Kr97}. 
\vskip 0.3cm

Recently, a new numerical analysis \cite{FMV} of the $4$-point
function of the scalar in two and three space dimensions
has been performed for different values of $\xi$.
It was based on a direct generation of samples of Lagrangian 
trajectories passing at time $t$ by $N$ fixed points.
Such samples allow a direct simulation of the PDF's $P_N(t;\unr,\unr')$ 
as well as direct simulations of the stationary correlation 
functions $F_N$. Let us explain how the latter were done
in \cite{FMV}. Upon setting $t_0=-\infty$ and $\theta(t_0)=0$
in Eq.\,\,(\ref{inhomo}), one obtains for even $N$ the relations
\qq
F_N(\unr)\ =\ \Big\langle\prod\limits_{n=1}^N
\int_{_{-\infty}}^{^t}\hspace{-0.2cm}ds\ f(s,\Nr_n(s))
\Big\rangle\ =\ 
\ \sum\limits_{\Pi}\Big\langle
\prod\limits_{\{n,m\}\in\Pi}\hspace{-0.3cm}
\CT(\Nr_n,\Nr_m)\,\Big\rangle
\label{ave}
\qqq
where the sum, resulting from the average over the Gaussian source,
is over the pairings $\Pi$ of the indices ${1,\dots, N}$ and where
\qq
\CT(\Nr_n,\Nr_m)\ =\ \int_{_{-\infty}}^{^t}\hspace{-0.3cm}ds
\ \m\CC({_{\Nr_n(s)-\Nr_m(s)}\over^L})
\label{times}
\qqq
is approximately equal to the time spent within distance $L$
by the two trajectories with the end-points $\Nr_n$ and $\Nr_m$.
The average of the product of such times may be simulated 
using the ensemble of trajectories passing through points $\Nr_n$. 
It is finite since the trajectories effectively separate as 
$(time)^{1\over2-\xi}$. Due to the limited dependence 
of trajectories on the initial conditions discussed above,
this remains true even if $\Nr_n$ and $\Nr_m$  tend to each other. 
The comparison of $F_N(\unr)$ calculated this way for different 
configurations of points $\Nr_n$ showed the dominance 
of the structure functions $S_N(r)$ by subleading terms 
of $F_N(\unr)$, as in the discussion at the end of Section 5. 
The resulting values of the anomalous $4$-point-structure-function 
exponent $2\zeta_2-\zeta_4$, presented in Fig.\,\,1 borrowed 
from \cite{FMV}, confirm the predictions of the perturbative 
zero-mode analysis around $\xi=0$ indicated in Fig.\,\,1 
by the broken-dotted line.

\begin{figure}
\centerline{\psfig{file=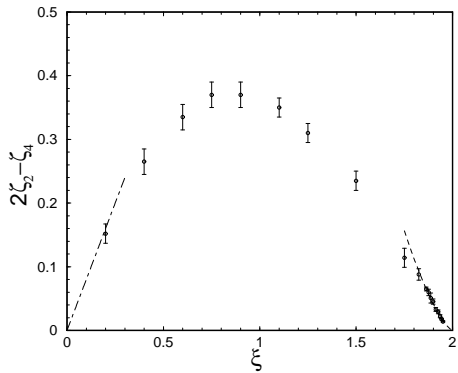,width=6.8cm,clip=}}
\caption{Anomalous exponent of the 4-point structure function
of the scalar as a function of $\xi$ \,
(U. Frisch, A. Mazzino and M. Vergassola, cond-mat/9802192)}
\end{figure}

\section{Conclusions}
\vskip -0.15cm

We have exhibited a mechanism behind intermittency
of the scalar structure functions in the Kraichnan model
of passive advection. The important point is that the
anomalous scaling originated from a discrete series 
of solutions of the homogeneous Hopf equations without 
the forcing and diffusion terms. This, and the relation 
of such solutions to the effective fuzzyness 
of Lagrangian trajectories at high Reynolds numbers
promise to be the stories that persist in other intermittent
hydrodynamical systems. In particular, it has been
indicated recently \cite{Lvov} how discrete solutions 
of the unforced Hopf equations may naturally give rise 
to a multiscaling picture of anomalous exponents in 
the Navier-Stokes case.


\vskip -0.33cm




\begin{thebibliography}{99}

\bibitem{Anton} R. Antonia, E. Hopfinger, Y. Gagne and F. Anselmet:
{\it Temperature Structure Functions in Turbulent Shear
Flows}. Phys. Rev. {\bf A 30} (1984), 2704-2707

\bibitem{Kr68} R.H. Kraichnan:
{\it Small-Scale Structure of a Scalar
Field Convected by Turbulence}.
Phys. Fluids {\bf 11} (1968), 945-963

\bibitem{Kr94} R. H. Kraichnan: {\it Anomalous Scaling 
of a Randomly Advected Passive Scalar}. Phys. Rev. Lett. 
{\bf 72} (1994), 1016-1019

\bibitem{GK} K. Gaw\c{e}dzki and A. Kupiainen:
{\it Anomalous Scaling of the Passive Scalar}.
Phys. Rev. Lett., {\bf 75} (1995), 3834-3837

\bibitem{CFKL} M. Chertkov, G. Falkovich, I. Kolokolov 
and V. Lebedev: {\it Normal and Anomalous Scaling 
of the Fourth-Order Correlation Function of a Randomly 
Advected Scalar}. Phys. Rev. {\bf E 52} (1995), 4924-4941

\bibitem{SS} B. Shraiman and E. Siggia: 
{\it Anomalous Scaling of a Passive Scalar
in Turbulent Flow}. C.R. Acad. Sci. {\bf 321} (1995), 279-284

\bibitem{BGK}D. Bernard, K. Gaw\c{e}dzki and A. Kupiainen:
{\it Anomalous Scaling of the N-Point Functions of a Passive 
Scalar}. Phys. Rev. {\bf E 54} (1996), 2564-2572

\bibitem{CF} M. Chertkov and G. Falkovich:
{\it Anomalous Scaling Exponents of a White-Advected
Passive Scalar}. Phys. Rev Lett. {\bf 76} (1996), 
2706-2709

\bibitem{SS2} B. Shraiman and E. Siggia: {\it Symmetry
and Scaling of Turbulent Mixing}. Phys. Rev. Lett.
{\bf 77} (1996), 2463-2466

\bibitem{GLPP} O. Gat, V. S. L'vov and I. Procaccia:
{\it Perturbative and Non-Perturbative Analysis of the 3'rd
Order Zero Modes}. Phys. Rev. {\bf E 56} (1997), 406-416

\bibitem{BCFL} E. Balkovsky, G. Falkovich and V. Lebedev:
{\it Three-Point Correlation Function of a Scalar Mixed
by an Almost Smooth Random Velocity Field}. Phys. Rev. {\bf E 55}
(1997), R4881-R4884

\bibitem{slow} D. Bernard, K. Gaw\c{e}dzki and A. Kupiainen:
{\it Slow Modes in Passive Advection}. 
cond-mat/9706035

\bibitem{KYC} R. H. Kraichnan, V. Yakhot and S. Chen:
{\it Scaling Relations for a Randomly Advected Passive
Scalar Field}. Phys. Rev. Lett. {\bf 75} (1995), 240-243

\bibitem{FGLP} A. L. Fairhall, B Galanti, V. S. L'vov and
I. Procaccia: {\it Direct Numerical Simulations
of the Kraichnan Model: Scaling Exponents and Fusion Rules}. 
Phys. Rev. Lett. {\bf 79} (1997), 4166-4169

\bibitem{Kr97} R. H. Kraichnan: {\it Passive Scalar: Scaling 
Exponents and Realizability}. Phys. Rev. Lett. {\bf 78}
(1979), 4922-4925

\bibitem{FMV} U. Frisch, A. Mazzino and M. Vergassola:
{\it Intermittency in Passive Scalar Advection}. cond-mat/9802192

\bibitem{Lvov} V. I. Belinicher, V. S. L'vov and I. Procaccia: 
{\it Computing the Scaling Exponents in Fluid
Turbulence from First Principles: Demonstration
of Multi-scaling}. chao-dyn/9708004


\end{thebibliography}
\end{document}